\documentstyle{l-aa}
\input{psfig.tex}
\begin{document}
                        
\thesaurus{03(11.02.1; 13.25.3) }

\def\a{\alpha}
\def\b{\beta}
\def\g{\gamma}
\def\G{\Gamma}
\def\d{\delta}
\def\D{\Delta}
\def\l{\lambda}
\def\t{\theta}
\def\m{\mu}
\def\n{\nu}
\def\r{\rho}
\def\s{\sigma}
\def\refitem{\par\parskip 0pt\noindent\hangindent 20pt}
\def\etal{{et\thinspace al.}\ }
\def\eg{{\it e.g.}\ }
\def\ie{{\it i.e.}\ }
\def\deg{{$^o$}}

\title{The blazar PKS 0528+134: new results from {\it Beppo}SAX observations}

\author{G. Ghisellini\inst{1}, L. Costamante \inst{1,2}, G. Tagliaferri\inst{1}, 
L. Maraschi\inst{3}, 
A. Celotti\inst{4}, 
G. Fossati\inst{5}, 
E. Pian\inst{6}, 
A. Comastri\inst{7}, 
G. De Francesco\inst{8}, L. Lanteri\inst{8}, 
C.M. Raiteri\inst{8}, G. Sobrito\inst{8}, M. Villata\inst{8}, 
% {\it S. Giarrusso\inst{9}, B. Sacco\inst{9}, }
I.S. Glass\inst{9},
P. Grandi\inst{10},
P. Padovani\inst{11},
% {\it G.G.C. Palumbo\inst{13},}
C. Perola\inst{12},
A. Treves\inst{13}
}

\offprints{G. Ghisellini}

\institute{
{Osservatorio Astronomico di Brera, V. Bianchi 46, I-23807 Merate, Italy}
\and{Univ. di Milano, V. Celoria 16, I-20100 Milano, Italy}
\and{Osservatorio Astronomico di Brera, V. Brera 28, I-20100 Milano, Italy}
\and{S.I.S.S.A., V. Beirut 2-4, I-34014, Trieste, Italy}
\and{CASS/UCSD, La Jolla, CA 92093-0424, USA}
\and{TESRE, Bologna, Italy}
\and{Osservatorio Astronomico di Bologna, V. Zamboni 33, I-40126 Bologna, Italy}
\and{Osservatorio Astronomico di Torino, Strada Osservatorio 20, I-10025 Pino Torinese, Italy}
% \and{IFCAI/CNR, Palermo, Italy}
\and{South African Astron. Observ.}
\and{IAS/CNR, Frascati, Roma, Italy}
\and{ESA--STScI, 3700 San Martin Drive, Baltimore, MD 21218, USA}
% \and{Univ. of Bologna, Bologna, Italy}
\and{Univ. of Rome III, Italy}
\and{Univ. of Como, Como, Italy}
}
\date{Received  ; accepted }
   
\maketitle

\markboth{Ghisellini et al., {\it Beppo}SAX observations of PKS 0528+134}{}

\begin{abstract}
{\it Beppo}SAX observed 8 times the $\gamma$--ray bright blazar 
PKS 0528+134 in
Feb. and Mar. 1997, during a multiwavelength campaign involving EGRET,
onboard CGRO, and ground based telescopes.  
The source was in its
faintest X--ray state observed so far, with an unabsorbed [2--10] keV
flux of $2.5\times 10^{-12}$ erg cm$^{-2}$ s$^{-1}$.  
The spectrum can be fitted with a power law, with energy index 
$\alpha_X=0.49 \pm 0.07$ between 0.1 and 10 keV.  
The high energy instrument PDS detected a flux in the 15--100 keV band 
which appears disconnected  from the lower X--ray energies, most likely 
due to a contaminating source.   
% The comparison  of
% this low state with previous higher states of the source shows
% an indication that the X--ray spectrum hardens and the $\gamma$--ray
% spectrum steepens when the source is fainter.  
We consider our findings in the context of the overall spectral energy 
distribution and discuss their implications for synchrotron and inverse
Compton models.

\keywords{X-rays: galaxies -- galaxies: active -- Individual: PKS 0528+134}
\end{abstract}

\section{Introduction}
Blazars received an unexpected boost of interest since the discovery,
driven by EGRET (and confirmed by OSSE and COMPTEL) that they emit 
the bulk of their radiative power at $\gamma$--ray energies.  
We are now finding spectral differences among blazars which have 
already been detected by EGRET.  
BL Lac objects seem to be characterized by a
$\gamma$--ray luminosity, $L_\gamma$, which is comparable to 
the luminosity emitted in the rest of the spectrum, while $L_\gamma$ 
tends to dominate in blazars with visible broad emission lines. 
At the same time, the $\gamma$--ray dominance correlates with the overall 
spectral energy distribution (SED) (Comastri et al. 1997, Fossati et al. 1998).  
All blazar SED are
in fact characterized by two broad emission peaks, believed to be
produced by the synchrotron and the inverse Compton processes.  
The location of these peaks (their frequency and relative flux) is a
strong diagnostic tool to discriminate among theoretical models and to
find the intrinsic physical parameters of the emitting region (see
e.g. Ghisellini, Maraschi \& Dondi 1996; Sikora et al. 1997; Dermer, 
Sturner \& Schlikeiser 1997).  
Examining the SED of $\gamma-$ray detected blazars, one finds
that both peaks are located at lower frequencies in more
powerful, and more $\gamma$--ray dominated, sources (Fossati et
al. 1998; Ghisellini et al. 1998).

PKS 0528+134, a very powerful blazar, is no exception: its SED shows
that the high energy peak is located in the 10-100 MeV range,
and the low energy peak between the far IR and the optical bands.  
For this source, the X--ray band falls near the minimum between 
the two peaks. 
Previous X--ray observations by the ASCA satellite
showed a flat ($\alpha<1$, with $F_\nu\propto \nu^{-\alpha}$) spectrum,
possibly connecting with the flux in the MeV--GeV band
(Sambruna et al. 1997).  
The uncertainty comes from the
paucity of simultaneous observations in both bands, and from the
relatively narrow energy windows of the used X--ray detectors.  
Indeed, if one could
join the X--ray and the $\gamma$--ray emission with a smooth curve
and, more importantly, simultaneous variability were detected,
then it would be possible to confidently argue that the
emission in both bands is produced by the same mechanism and in 
the same region.

Furthermore, the fact that the X--ray emission of this source is
in a minimum of the SED is of particular interest.  In fact, the
possible reprocessing of high energy photons, via the creation of
electron-positron pairs, leads inevitably to the production of
radiation, mainly in the X--ray band. 
It is therefore crucial to determine  how deep this minimum is: 
a deep minimum would be a signature that this process {\it does not} 
occur (Ghisellini \& Madau 1996). 

For these reasons we have undertaken a program to observe PKS 0528+134 
with the X--ray broad band {\it Beppo}SAX satellite
simultaneously with CGRO (in particular with the EGRET instrument) 
and other, ground-based, telescopes.  
In this paper we mainly present the {\it Beppo}SAX
observations, compare them with previous X--ray observations, and
construct a simultaneous SED with the optical and EGRET data, briefly
discussing our findings.

\begin{table*}
\begin{center}
\begin{tabular}{l l l r c r c r c}
\multicolumn{9}{c}{\bf TABLE 1: {\it Beppo}SAX observation log} \\
\hline
\hline
Date &Start$^1$ &End$^1$ &LECS$^2$ &net cts/s  &MECS$^2$ &net cts/s &PDS$^2$  &net cts/s $^3$ \\
\hline
Feb 21  &02:17:38 &09:08:06 &6322 &0.015$\pm$0.007 &14487 &0.042$\pm$0.002  &6919 &0.22$\pm$0.08  \\
Feb 22  &23:29:16 &06:48:06 &5046 &0.014$\pm$0.007 &13365 &0.045$\pm$0.003  &5925 &0.24$\pm$0.08  \\ 
Feb 27  &14:38:39 &19:18:11 &3945 &0.011$\pm$0.010 &6917  &0.041$\pm$0.004  &1622 &0.28$\pm$0.17  \\
Mar 1   &00:52:14 &07:43:06 &4261 &0.012$\pm$0.010 &13547 &0.043$\pm$0.003  &6344 &0.29$\pm$0.08  \\
Mar 3   &01:38:02 &08:28:06 &5053 &0.015$\pm$0.007 &14241 &0.039$\pm$0.002  &6783 &0.16$\pm$0.08  \\
Mar 4-5 &21:53:08 &04:28:05 &2793 &0.015$\pm$0.010 &11254 &0.043$\pm$0.003  &4914 &0.11$\pm$0.09  \\
Mar 6-7 &21:03:14 &03:53:05 &3083 &0.012$\pm$0.010 &12762 &0.041$\pm$0.003  &5587 &0.26$\pm$0.09  \\
Mar 11  &00:17:06 &06:06:05 &2497 &0.012$\pm$0.011 &11490 &0.037$\pm$0.003  &5374 &0.29$\pm$0.09  \\
\hline
Total   &         &        &32998 &0.014$\pm$0.001 &98063 &0.041$\pm$0.001 &43468 &0.23$\pm$0.03  \\
\hline
\hline
\multicolumn{9}{l}{$^1$ UT Time; $^2$ Exposure time in seconds; 
$^3$ values in the entire PDS band } \\
\end{tabular}
\end{center}
\end{table*}

\section{PKS 0528+134}

PKS 0528+134 ($z=2.07$) is one of the most distant quasar detected by
EGRET in the $\gamma$--ray band.  
Located in the galactic anticenter region, it is heavily absorbed, 
although the estimates of $A_V$ are very uncertain (ranging from 2.3 to 5).  
Thus PKS 0528+134 is faint in
the optical, with a typical average magnitude of 19.5 in the $V$ band.
It is a strong and flat radio source, with detected superluminal
motion ($\beta_{app}\sim$4--6 for $H_0=100$ km s$^{-1}$
Mpc$^{-1}$, Pohl et al. 1996; Britzen, Witzel \& Krichbaum 1996).
Close to Geminga and the Crab, it was frequently observed by EGRET,
and seen flaring in March 1993, when the $\gamma$--ray flux was about 6 
times brighter than average (Hunter et al. 1993; Mukherjee et al. 1996).  
Flux changes in the $\gamma$--band are accompanied by
spectral variations in the sense that the spectrum is harder when
brighter (Sambruna et al. 1997; Mukherjee et al. 1997a, 1997b).
It is also one of the few extragalactic sources detected by 
COMPTEL (Collmar et al. 1997; B\"ottcher \& Collmar 1998).  
The combined COMPTEL and EGRET observations show that the peak of the high 
energy emission occurs in the MeV band, at least in the flaring state.

\section{{\it Beppo}SAX observations}

Between Feb. 21 and March 11, 1997, {\it Beppo}SAX observed the source 
8 times with the narrow field instruments 
LECS (Low Energy Concentrator Spectrometer; energy range 0.1--10 keV), 
MECS (Medium Energy Concentrator Spectrometer; energy range 1.3--10 keV) 
and PDS (Phoswich Detector System; energy range 15--200 keV)
(Boella et al. 1997).

In Table 1 we report the journal of observations, with the corresponding 
exposure times and derived net count rates.

\begin{figure}
\psfig{file=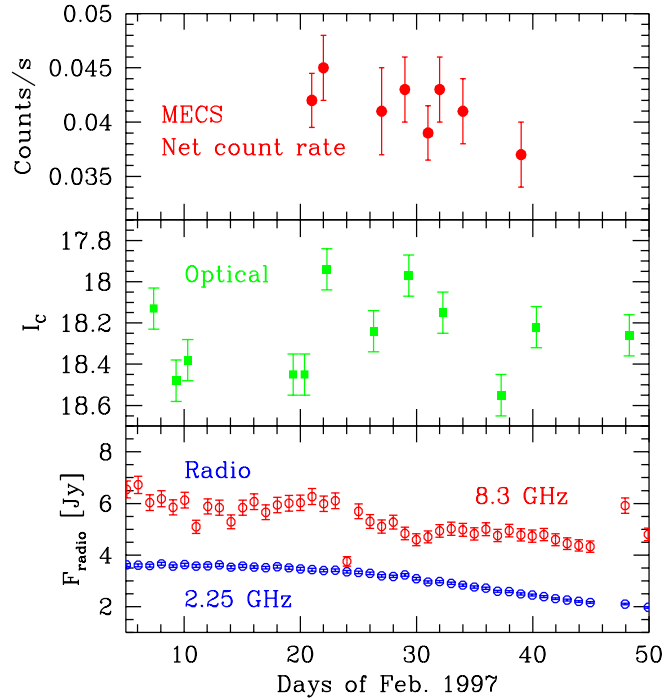,width=9.5truecm,height=10truecm}% lc_opt_sax
\caption[h]{The top panel shows the MECS light curve, 
corresponding to the 8 observations.  In the middle panel we report
the optical light curve in the same period, in the $I_C$ filter.  The
bottom panel shows the light curve at 8.3 and 2.25 GHz, by the Green
Bank monitoring campaign.}
\end{figure}

\section{Results}

\subsection{Light curves}

We have extracted light curves from the data of the three MECS units 
(merged together) using XSELECT and considering a circle of radius 
$4^\prime$ centered around the target.  
In Fig. 1 we show the total MECS count rates obtained with the XIMAGE
program corresponding to the 8 observations.  
This light curve is consistent with a constant. 
We can set an upper limit on the amplitude of possible variations of 
$\sim$40\%.
%(probability of 60\% to obtain a 
%larger value of $\chi^2$, when fitting a constant). 
No significant variations are present within the single observations either.

Simultaneously with the {\it Beppo}SAX observations, the source was observed
in the optical, with the REOSC 1.05 meter telescope of Pino Torinese
(see Villata et al. 1997 for a description of the instrument and the
data analysis procedure), and in the radio band with the Green
Bank Interferometer (GBI).  
As shown by the middle and lower panels of Fig. 1, the source was 
varying significantly in the optical (with a probability $>99.9\%$) 
and at 8.3 GHz.

We also observed the source in the near infrared with the 1.9 m telescope
at Sutherland using the MkIII photometer, obtaining $K=14.76\pm 0.23$ 
on Mar 2, 1998 (JD=2450510), using a 12" diameter aperture.

\subsection{Spectral fitting}
 
\subsubsection{LECS and MECS}

Spectra of the LECS and MECS detectors have been extracted using
XSELECT and a circle of radius 4$^\prime$ around the target for both
the instruments (given the weak flux of the source and the heavy absorption
under 0.7 keV, we have not used 
the $8.5^\prime$ radius recommended for the LECS for brighter sources,
to minimize the effects of the background, especially at low energies).  
As suggested by the Science Data Center of {\it Beppo}SAX (SDC), 
after checking that the background was constant during the observations,
we have subtracted the background using the blank sky fields selecting 
the same detector regions where the target is located.
We used the response matrices released by the SDC in September 1997.
The LECS and MECS spectra have been jointly fit after allowing for a constant
rescaling factor of 0.9 for the LECS data 
(this factor accounts for uncertainties in
the inter--calibration of the instruments; Fiore, priv. comm.).

Fitting the 8 observations separately with a single power law and free
hydrogen column density, $N_{\rm H}$, gives consistent values, within
the (rather large) uncertainties.
We therefore decided to combine the 8 spectra, in order to increase 
the signal to noise ratio.  
A single power law fit to the LECS+MECS dataset (see Fig. 2) yields 
an energy spectral index $\alpha_{x}=0.48\pm 0.11$ with 
$N_{\rm H}=(5.0\pm 1.9)\times 10^{21}$ cm$^{-2}$ (unless otherwise 
indicated, the errors from the fits are at 90\% confidence level, 
for two parameters of interest).
No spectral feature is required by
the fit and in particular an upper limit of $EW<200$ eV can be set to
the equivalent width (in the source frame) of the 6.4 keV fluorescence
$K_\alpha$ iron line (see \S 5.1 for more discussion). 
The results of the spectral fitting are reported in Table 2.

\begin{figure}
\vskip -3 true cm
\psfig{file=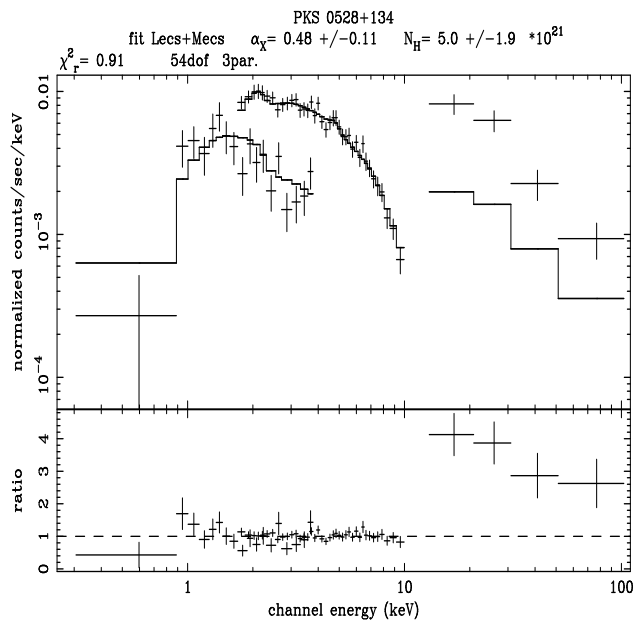,width=8.5truecm,height=15truecm} % sax fit 
\vskip -3.5 true cm
\caption[h]{Fit to the LECS and MECS {\it Beppo}SAX spectrum.
PDS data are not used for the fit, and appear to lie above the
extrapolation of the power law from lower energies. 
}
\end{figure}

\begin{table*}
\begin{center}
\begin{tabular}{l l l l l l l }
\multicolumn{7}{c}{\bf TABLE 2: Fits to {\it Beppo}SAX Data: single power law} \\
\hline
\hline
% \multicolumn{7}{l}{$^1$ UT Time, $^2$ Exposure time in seconds.  }\\
$\alpha$      &$N_{\rm H}$              &$F_{[2-10~{\rm keV}]}$  &$\chi^2_r$ &d.o.f.  &$F_{\rm 1~keV}$ &Notes \\       % Chi2 con PDS     
              &$10^{21}$cm$^{-2}$ &erg/cm$^{2}$/s &   &    &$\mu Jy$ & \\      
\hline
0.48$\pm$0.11 &5.0$\pm$1.9        &2.53e-12       &0.91       &54 &0.29$\pm$0.05 &LECS$+$MECS, free $N_{\rm H}$  \\
0.49$\pm$0.07 &5.3                &2.54e-12       &0.91       &55 &0.30$\pm$0.03 &LECS$+$MECS, fixed $N_{\rm H}$ \\ % 1.04 (59-2) 
0.37$\pm$0.11 &3.9$\pm$1.6        &2.59e-12       &1.71       &58 &0.24$\pm$0.04 &LECS$+$MECS$+$PDS, free $N_{\rm H}$ \\
0.42$\pm$0.06 &5.3                &2.60e-12       &1.74       &59 &0.27$\pm$0.03 &LECS$+$MECS$+$PDS, fixed $N_{\rm H}$ \\
\hline
\hline
\end{tabular}
\end{center}
\end{table*}

\begin{table*}
\begin{center}
\begin{tabular}{l l l l l l}
\multicolumn{6}{c}{\bf TABLE 3: Single power law fits to ASCA data with fixed $N_{\rm H}$}\\
\hline
\hline
$\alpha$      &$F_{[2-10~{\rm keV}]}$          &$\chi^2_r$ &d.o.f.   &$F_{\rm 1~keV}$  &Date \\       % Chi2 con PDS     
              &erg cm$^{-2}$s$^{-1}$ &           &    &$\mu Jy$    &    \\      
\hline
0.66$\pm$0.06 &3.77e-12       &0.89       &93  &0.57$\pm$0.05  &28 Aug 1994   \\
0.71$\pm$0.04 &9.55e-12       &1.21       &100 &1.57$\pm$0.07  &7 Mar 1995   \\
0.71$\pm$0.04 &10.62e-12      &1.18       &69  &1.76$\pm$0.07  &14 Mar 1995   \\
0.72$\pm$0.03 &13.80e-12      &0.91       &140 &2.32$\pm$0.07  &19 Mar 1995   \\
\hline
\end{tabular}

\begin{tabular}{l l l l l l l l }
\multicolumn{8}{c}{\bf Fits to the summed ASCA data of  March 95} \\
\hline
$\alpha_1$ &$\alpha_2$ &$E_{break}$   &$N_{\rm H}$              &$F_{[2-10~{\rm keV}]}$ &$\chi^2_r$ &d.o.f.  &Notes \\ 
           &           &keV           &$10^{21}$cm$^{-2}$ &erg/cm$^{2}$/s &           &        & \\      
\hline
0.72$\pm$0.04 &       -       &       -          &5.3$\pm0.3$ &1.15e-11  &1.13  &207  &free $N_{\rm H}$, single power law  \\
0.00$\pm$0.16 &$0.65\pm$0.03  &$1.47\pm0.10$     &3.9 fix     &1.15e-11  &1.08  &206  &fixed $N_{\rm H}$, broken power law \\
\hline
\hline
\end{tabular}
\end{center}
\end{table*}

As mentioned in the Introduction, the value of the $N_{\rm H}$ column
density is uncertain, yet its determination is important, since
the knowledge of the corresponding optical extinction allows to
determine the location of the synchrotron peak.  {\it Beppo}SAX data, alone,
cannot well constrain $N_{\rm H}$, due to the faint level of the source.  
PKS 0528+134  was instead much brighter during the
ASCA observations of March 1995, allowing a better
determination of the $N_{\rm H}$ column [$N_{\rm H}=5.3(\pm
0.3)\times 10^{21}$ cm$^{-2}$], as described in \S 5.1.

Since the {\it Beppo}SAX measurement of the $N_{\rm H}$ column is consistent
with the value found by ASCA, we 
fixed the $N_{\rm H}$ at the latter value.  
In this case the LECS+MECS data can be modeled by a single 
power law with an energy index $\alpha_x=0.49\pm 0.07$
(90\% confidence level for one parameter of interest).
% The same data can be fitted with a broken power law model (see Table 2),
% but without a significant improvement of the fit.

\subsubsection{PDS}

Source visibility windows were selected following the criteria of no earth
occultation and high voltages stability during the exposures.  
In addition, the observations closest to the South Atlantic anomaly were
discarded from the analysis.  From collimators positions 
the ON and OFF time windows were also 
created and merged with the source visibility window to create the 
final time windows on which the source+background and background 
spectra were accumulated for each of the four PDS units, using
the XAS software package.  
A filtering for the temperature and energy dependence of the pulse rise 
time was used. 

The source was significantly detected up to $\sim$90 keV.
For each pointing, the grouped spectra from the four units were coadded.  
The net count rates are reported in Table 1.  

To increase the S/N, we have averaged the eight PDS spectra
and binned the resulting spectrum in 4 energy intervals. 
For the fitting, we used the response matrix available at SDC,
keeping the rescaling factor with respect to the MECS constant and equal 
to 0.85 (this factor allows for uncertainties in the 
inter--calibration of the instruments).

As shown in Fig. 2, the PDS data points lie above the model fit for
the LECS and MECS datasets.  In particular, including the PDS data and
using a single power law fit with $N_{\rm H}$ fixed to the value found
by ASCA, we obtain $\alpha_{x}=0.42\pm 0.06$, with the PDS points
still lying above the fit (see Table 2 for details).

The same three datasets (LECS+MECS+PDS)  were also fitted with a
broken power law model with $\alpha_1=0.48\pm 0.06$ and
$\alpha_2=-0.15\pm 0.14$ and break energy $E_{break} \sim 9^{+3}_{-2}$
keV, leading to a $\chi^2_r=1.34$ for 57 d.o.f.  This model
significantly improves the fit, according to the $F$--test ($>99.5\%$),
but the $\chi^2$--value yields a probability less than 5\% that the model
is correct.
A power law fit to the PDS data alone gives a spectral index 
$\alpha_{x}=0.82\pm 0.35$.

As can be seen, the high energy flux detected by the PDS shows 
a clear discontinuity with respect to the LECS/MECS data.  
A possible explanation can be that the PDS data are contaminated by the
presence of another source in its field of view.  
The PDS field of view is $\sim$1.3 degree, with no imaging capability.
The response matrix of this instrument is triangular, with a flat top 
of 3$^\prime$ and a reduction of a factor 2 in sensitivity at 38$^\prime$
from the center (Frontera et al. 1997).
The MECS have a field of view of radius $30^{\prime}$, and there are 
no sources other than PKS 0528+134 detected at more than 3$\sigma$.

We checked the ROSAT PSPC WGA catalog (White, Giommi \& Angelini, 1994) 
for sources at an angular distance $<1$ degree from PKS 0528+134.  
There are two unidentified sources 
(1WGA J0532.9+134, at a distance of 31 arcmin from PKS 0528+134, 
and 1WGA J0533+135, at 41 arcmin) with a count rate
in the ROSAT band comparable to that of PKS 0528+134. 
No sources were instead found in the EXOSAT and ASCA public archives.  
Given the low galactic latitude of PKS 0528+134 and the large offset
of the PDS data points with respect to the extrapolation of the MECS spectrum,
we think that the PDS data might be contaminated by other sources.
For this reason, in the discussion of the SED of PKS 0528+134,
we will not consider the PDS spectrum.

% In conclusion, we cannot either exclude or confirm that other sources 
% contribute to the relatively strong PDS flux of PKS 0528+134. 
% Given the large spectral ``discontinuity" implied by those data we will 
% provisionally assume that they do not refer to PKS 0528+134.

\section{Comparison with previous observations}

\subsection{ASCA observations}

As already mentioned, a long standing problem with this source
is the determination of its optical extinction and the column 
$N_{\rm H}$ along the line of sight.  
The value that we found with the
{\it Beppo}SAX observations is in agreement with the sum of the estimated
absorption caused by the column of neutral galactic hydrogen 
[$N_{\rm H}=(2.6\pm 0.1)\times 10^{21}$ cm$^{-2}$], plus that due to 
the outer edge of the molecular cloud Barnard 30 in the $\lambda$ 
Orion rings of clouds, [$N_{\rm H}\sim 1.3\times 10^{21}$ cm$^{-2}$]
(Sambruna et al. 1997).

\begin{figure}
\vskip -4 true cm
\psfig{file=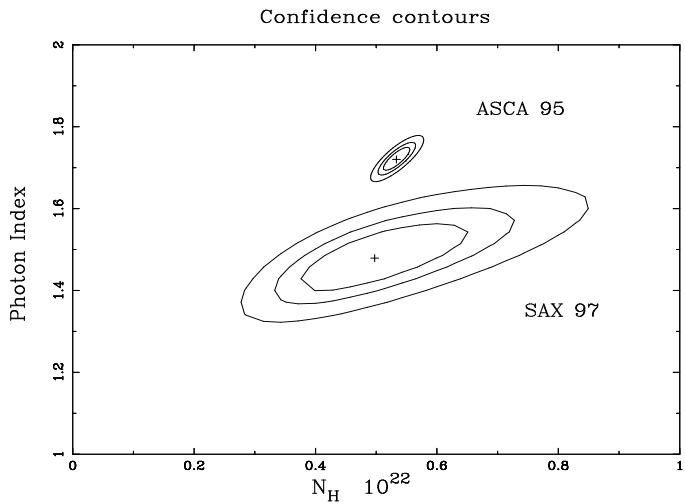,width=9truecm,height=11truecm}% contour 97
\caption[h]{Confidence contours (68\%, 90\% and 99\%) for fits with a
single power law plus absorption for the sum of the three August 1995
ASCA observations and the sum of the eight {\it Beppo}SAX observations.}
\end{figure}

In March 1995, ASCA observed PKS 0528+134 three times, within 2 weeks,
during an active state of the source.  
The last of these observations was analyzed by Sambruna et al. (1997), 
finding a best fit value of $N_{\rm H}= (5.57\pm 0.4)\times 10^{21}$
cm$^{-2}$, in good agreement with the value found in our {\it Beppo}SAX
observations.  
The entire set of the three observations was instead
analyzed by Kubo et al. (1998), but assuming a column density
fixed at the neutral galactic hydrogen value 
[$N_{\rm H}=(2.6\pm 0.1)\times 10^{21}$ cm$^{-2}$], 
without considering the extra absorption due to the Barnard cloud.

We therefore decided to re--analyze the full ASCA data set.
We first fitted the 3 observations separately, finding that,
despite a flux change of a factor 1.5, the three best fit values of
the spectral index were nearly equal: with free $N_{\rm H}$ 
the values were within the 68\% confidence uncertainties.  
We then fitted the sum of the 3 ASCA observations and the increased
signal--to--noise ratio allowed to obtain the best measure of the 
column so far: $N_{\rm H}=(5.3\pm 0.3)\times 10^{21}$ cm$^{-2}$
(see Table 3 for the fit parameters).

% (QUA IL PROBLEMA BROKEN POWERLAW PER ASCA ?  F--test da` preferibile
% la doppia powerlaw a $\sim 99\%$, partendo da 0.7 keV coi SIS, ma dato
% che ci sono ancora residui problemi di calibrazione per sorgenti
% diverse dalla CRAB, e piccole differenze a bassa energia in confronto
% al LECS.....  non ci crediamo tanto. Vedi ASCA GOF Calibration
% uncertainties $http://heasarc.gsfc.nasa.gov/docs/asca/cal_probs.html)$.

In Fig. 3 we show the confidence contours in the photon index--column
plane, for the sum of the August 1995 ASCA observations and the sum of
the 8 {\it Beppo}SAX observations.

We also searched for the possible presence of an Fe
$K_\alpha$ line at 6.4 keV, finding an upper limit to its
equivalent width of 67 eV in the summed 1995 spectrum (calculated in
the rest frame of the source).  
This limit is only marginally consistent with the value of 
$EW=119\pm58$ eV reported by Reeves et al. (1997) analyzing the 1994
ASCA spectrum, when the source was fainter.  
% This discrepancy can possibly be due to the fact that Reeves et al. 
% (1997) assumed a galactic value of $N_{\rm H}=2.3\times 10^{21}$ 
% cm$^{-2}$ (i.e. they ignore the additional galactic component) and added 
% $N_{\rm H}=4.2\times 10^{22}$ cm$^{-2}$ at the blazar location.

\subsection{Historical X--ray and $\gamma$--ray light curves}

In order to understand the relationship between X--ray and
$\gamma-$ray emission, we compared both the light curves in the last 7
years and the spectral indices behaviour in these two bands. 

In Fig. 4 (upper panel) we show the light curves of all the
observations in the X--ray band after 1991. 
For consistency, we have re--analyzed all these data (taken from the 
archives) with the same $N_{\rm H}=5.3\times 10^{21}$ cm$^{-2}$. 
In 1995 the source varied by 50\% in 2 weeks.  
Between the 1995 ASCA and our 1997 {\it Beppo}SAX observations, when the 
source was at its faintest historical level, the flux decreased by
a factor $\sim$7.  
The lower panel shows the light curve of the $\gamma$--ray flux, 
as observed by EGRET (Mukherjee et al. 1996, 1997a, 1997b).  
In $\gamma$--rays the source was observed more often, with a variability 
of a factor 13 in 2 months, and a factor $\sim$2 in 2 days during 
the 1993 flare (see Mukherjee et al. 1996).

The lack of simultaneity of previous observations and the paucity
of X--ray data do not allow to establish any (or absence of)
correlation.  
It should be however noticed that for the three epochs of simultaneous 
observations (Aug. 1994, Mar. 1995 and Feb.--Mar. 1997), 
the X--ray and $\gamma$--ray flux levels follow the same trend.

\begin{figure}
% \vskip -1 true cm
\psfig{file=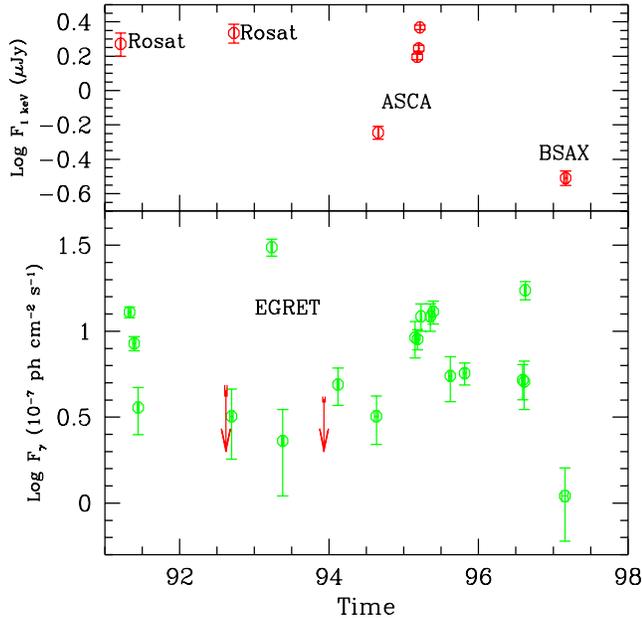,width=9.5truecm,height=10truecm} % lc_log_x_g
\vskip -1.5 true cm
\caption[h]{Historical light curves in the X--ray and $\gamma$--ray
bands, after 1991.  All X--ray data have been reanalysed assuming a
fixed $N_{\rm H}=5.3\times 10^{21}$ cm$^{-2}$. }
\end{figure}

\subsection{Flux--spectral index correlations?}

In Fig. 5 (upper panel) we report the X--ray spectral indices
determined by {\it Beppo}SAX and ASCA data vs the 1 keV flux. 
While all ASCA points are consistent with the same spectral index, 
the spectral shape determined by {\it Beppo}SAX observations is 
significantly flatter (see also Fig. 3), and corresponds to the fainter flux.

%There is an indication of a flattening of the slope when the source is
%fainter, just the opposite of what (probably) happens in the $\gamma$--ray band 

In the lower panel of Fig. 5 we show the $\gamma$--ray spectral index
as a function of the $\gamma$--ray flux
(data from Mukherjee et al. 1996, 1997a, 1997b, 1999). 
The large error bars of $\alpha_\gamma$, especially at faint fluxes, 
do not allow any conclusion 
about the presence of a trend: although the linear correlation coefficient
is $\sim$--0.6 (random probability=0.02), a $\chi^2$--test on a fit with a 
constant gives 15\% probability that the two quantities are not correlated.

Although we cannot draw any firm conclusion about the flux--spectral index
correlations, note that during the {\it Beppo}SAX observations the source,
in its faintest state, had the flattest spectral index.
This is an unusual behavior for blazars
(even if some other example exist, see  Ulrich,  Maraschi \&  Urry 1997), 
and likely to yield important information and/or constraints on 
the emission models, as discussed below.

% nevertheless the two sets of the ASCA and 
% {\it Beppo}SAX data are 
% significantly different in flux and spectral index.

% While in the $\gamma$--ray band the ``flatter when brighter" behavior is
% similar to what observed in other sources
% (see e.g. Ulrich,  Maraschi \&  Urry 1997), 
% the ``flatter when fainter" trend in the X--ray band is unusual
% (even if some other example exist, see  Ulrich,  Maraschi \&  Urry 1997), 
% and likely to yield important information and/or constraints on 
% the emission models.

\begin{figure}
\psfig{file=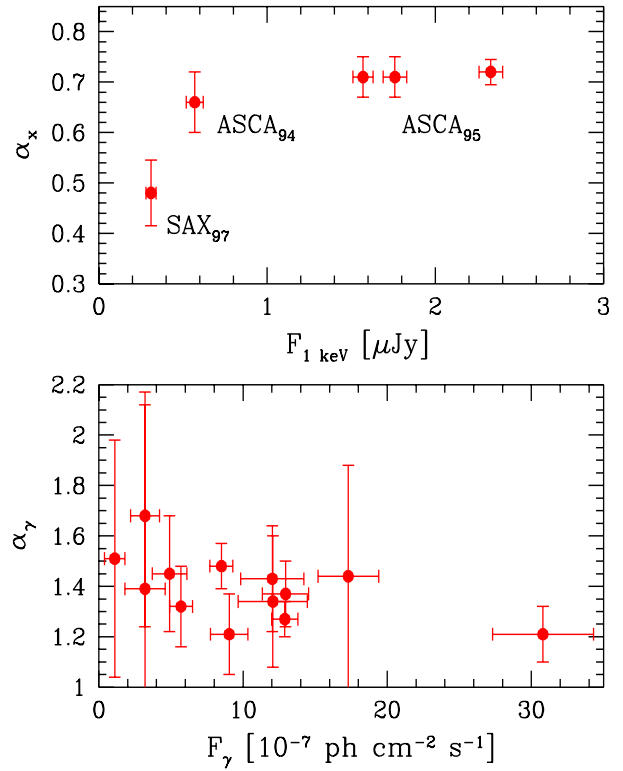,width=10truecm,height=11truecm} % al_f
\vskip -0.7 true cm
\caption[h]{
Spectral indices vs flux in the [2--10] keV and [0.05--1] GeV bands.  
All X--ray data have been reanalysed
assuming a fixed $N_{\rm H}=5.3\times 10^{21}$ cm$^{-2}$.
$\gamma$--ray data from Mukherjee et al. 1996, 1997a, 1997b, 1999.
}
\end{figure}

\section{The SED}

In Fig. 6 we show the overall spectral energy distributions of PKS
0528+134 at different epochs, as indicated by the labels.  
Data have been dereddened assuming $N_{\rm H}=5.3 \times 10^{21}$ cm$^{-2}$, 
corresponding to $A_V=3$.  
We have reported the minimum and maximum values
of the radio and optical data obtained during our campaign.

The not simultaneous data shown in Fig. 6 have been taken from
Wall \& Peacock 1985; Bloom et al. 1994; Edelson 1987; Reuter et al. 1997;
Rieke et al. 1982; Condon et al. 1977; Collmar et al. 1997;
McNaron-Brown et al., 1995 and Mukherjee et al. 1996.
We also show some simultaneous sets of data:
the 1991 Comptel and EGRET spectrum (Collmar et al. 1997);
the 1994 and 1995 X--ray (ASCA) and EGRET data (we also have an optical 
point during the 1994 campaign; data for these two campaigns are presented 
in Sambruna et al. 1997 and references therein);
our 1997 IR, optical, X--ray data together with the EGRET point from 
Mukherjee et al. 1997b and the radio data from the GBI archive (see Fig. 1).

As all other $\gamma$--ray bright blazars, also the SED of
PKS 0528+134 is characterized by two peaks, one between  the far IR 
and optical spectral bands, and the other at MeV energies.  
As discussed in Sambruna et al. (1997), the dereddened  (using $A_V=3$) 
optical spectrum is inverted ($\alpha_o=-0.18\pm 0.08$), 
and thus indicative of the presence of a `blue bump' component. 
This is not easy to reconcile with the very large and rapid optical 
variability. 
We also note that due to the presence of molecular clouds along 
the line of sight the gas to dust ratio may be anomalous and the
value of $A_V$ may be larger than the one adopted above.
Further spectral observations in the IR optical range could help 
clarifying this issue.

As can be seen, the MeV--GeV emission dominates the bolometric output 
by a large amount, reaching an (isotropic) luminosity in excess of 
10$^{49}$ erg s$^{-1}$ (with $H_0=50$ km s$^{-1}$ Mpc$^{-1}$ and $q_0=0.5$).

% Sources of data: {\it Not simultaneous}: 
% Wall \& Peacock 1985; Bloom et al. 1994; Edelson 1987; Reuter et al. 1997;
% Rieke et al. 1982; Condon et al. 1977; Collmar et al. 1997;
% McNaron-Brown et al., 1995; Mukherjee et al. 1996.
% {\it 1994 and 1995}: Sambruna et al. 1997 and references therein.
% {\it 1997}: this paper and Mukherjee et al. 1997.

\begin{figure*}
\vskip -0.7 true cm
\psfig{file=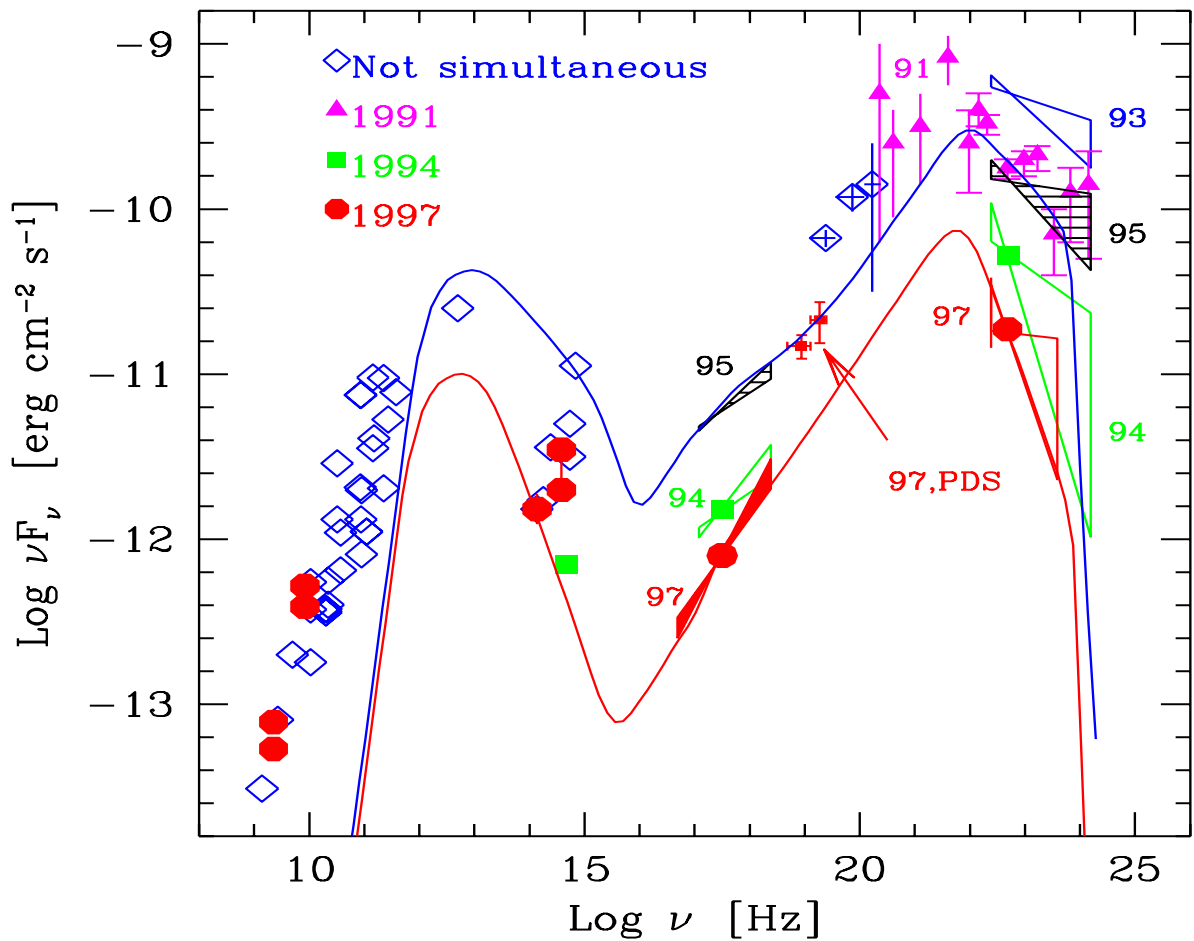,width=18truecm,height=15truecm}% SED
\vskip -1.5 true cm
\caption[h]{Overall spectral energy distribution of PKS 0528+134.
Simultaneous observations are indicated by the different
marks, as labeled. 
Other non--simultaneous data are taken from the literature
(see text).  
The solid line is a fit to the 1997 spectrum made with a
homogeneous, one--zone model, corresponding to a source with a size
$R=5\times 10^{16}$ cm, a magnetic field $B=8.6$ Gauss, in
which beaming corresponds to a Doppler factor $\delta=17$.
Relativistic electrons are injected continuously throughout the
source, at a rate corresponding to an intrinsic luminosity 
$L^\prime=3.7\times 10^{44}$ erg s$^{-1}$, with a power law spectrum 
$\propto\gamma^{-3.8}$ between $\gamma_1=200$ and $\gamma_2=3000$.  
In this case the EC component dominates the entire high energy band.
The fit to the 1995 spectrum is obtained adopting the same $R$,
$\delta$ and $B$, but with a larger injected power of
$L^\prime =1.8\times 10^{45}$ erg s$^{-1}$ and with a flatter injected 
electron spectrum ($\propto \gamma^{-2.7}$)
between $\gamma_1=200$ and $\gamma_2=4000$.  
In this case both the SSC and the EC components contribute to the 
low energy X--ray spectrum, while only the EC component contributes 
above a few tens of keV. 
Note that the shown PDS points lie above the extrapolation from
the spectrum in the LECS and MECS band.
Given the possibility of contaminating sources (see text), we do not
consider these points when applying the model.
}
\end{figure*}

\section{Discussion}

Among the models proposed to account for the blazar $\gamma$--ray emission,
the most accredited are those based on the Inverse Compton 
process, between the relativistic electrons
responsible for the synchrotron emission and seed photons
which can either be the synchrotron photons themselves 
[synchrotron--self Compton (SSC)] or photons produced externally to the
jet (EC) 
(Maraschi, Ghisellini \& Celotti 1992; Bloom \& Marscher 1993;
Sikora, Begelman, \& Rees 1994, Dermer, Schlickeiser \& Mastichiadis 1992;
Blandford \& Levinson 1995; Ghisellini \& Madau 1996).

In all cases synchrotron radiation is responsible for the low energy 
emission (up to the optical-UV band), and the inverse Compton process 
gives rise to the X-- to $\gamma$-ray spectrum.  
Seed photons of different origin may well coexist and their 
relative contribution in the same object may be different in different 
states and in different energy bands.
The presence of luminous emission lines and possibly of a blue bump indicates,
in the case of PKS 0528+134, that photons outside the jet
should play an important role in the IC process (EC scenario). 
For a detailed discussion of the fitting of the 1994 overall spectrum 
with SSC and EC models see Sambruna et al. (1997).

As shown in Fig. 6, the source shows correlated X--ray and $\gamma$--ray
flux variability. 
In particular, during the 1997 campaign, the source was at
its faintest level both in X-- and $\gamma$--rays.
In this occasion, the X--ray spectrum was significantly
flatter than in higher states.
This is in contrast with the ``flatter when brighter" behavior of
blazars (Ulrich, Maraschi \& Urry, 1997), and therefore likely
to give important constraints on the emission models.

Within the framework of the SSC and EC model we can envisage two
scenarios to explain the observed behavior:

\vskip 0.5 true cm

1) The X--rays are produced by $both$ the self--Compton and EC processes.  
Therefore there can be $two$ typical frequencies of
the seed photons: one corresponding to the peak of the synchrotron
emission (in the far IR), and one corresponding to the external radiation.  
In the case that the emission lines and blue bump
photons form the bulk of the external radiation, their typical
frequency, as observed in the comoving frame, is in the far UV.  
If the radiation energy densities of these two components are comparable
(within a factor 10), then the self--Compton emission would 
dominate at lower X--ray energies, while the EC spectrum 
would be entirely responsible for the emission in the $\gamma$--ray band.  
This is because the self--Compton spectrum is somewhat
steeper than the EC one produced by the same electrons, and because
the maximum self--Compton frequency is lower.  
Let us suppose now that the number of emitting electrons increases. 
In this case the synchrotron and the EC fluxes vary linearly, 
while the self--Compton one varies quadratically, making the SSC component 
to dominate the flux over a larger X--ray energy range.  
% This could account for the ``steeper when brighter" behavior of 
% PKS 0528+134 in X--rays.  
% To explain the opposite trend in $\gamma$--rays it
% is necessary to postulate, in addition, that the variation of
% the electron number is accompanied by a flattening of the high energy
% tail of their energy distribution, a behavior often invoked to 
% interpret blazar variability.

We have reproduced the 1997 ``low state" and the 1995 ``high state" 
spectra along these lines, with a homogeneous one--zone model, taking 
into account both the SSC and the EC contributions to the high
energy spectra. 
The model finds the equilibrium (steady state) electron distribution by 
balancing the injection of relativistic particles and their radiative cooling.
Photon--photon collisions and electron--positron pair production, 
Klein--Nishina effects on the scattering cross section and Coulomb 
collisions are taken into account (see Sambruna et al. 1997 
for further details). 
The input parameters are given in the caption of Fig. 6.  We ``fitted" both
spectra leaving the dimension and the beaming factor of the source  unchanged, 
as well the low energy cut-off of the injected electron distribution.  
We have instead increased by a factor 3 the injected power in the high state, 
also characterized by a slightly flatter injected electron distribution.  
The resulting spectra, shown in Fig. 6, reproduce quite 
well the different X and $\gamma$--ray slopes.
% We regard the discrepancy with the 1997 PDS spectrum as not
% significant, since the PDS could be contaminated by other sources.

\vskip 0.5 true cm

2) Another process that can in principle account for a flatter X--ray 
spectral index when the source is fainter, is e$^\pm$ pair production. 
When the source is in a high (and hard) $\gamma$--ray state, 
a small fraction of the high energy power could be converted in 
electron--positron pairs via $\gamma -\gamma$ interaction with  
X--ray photons. 
This process produces a steepening of the electron spectrum towards lower
energies, yielding a brighter and steeper IC spectrum in the X--ray band.
The main difficulty with this scenario is the fine tuning required in the 
amount of the high energy power that is absorbed and reprocessed, 
which must be of the order of a few per cent. 
If it is less, then pairs are not produced in a sufficient amount 
to contribute anywhere in the emitted spectrum.  
If it is higher, then they could
increase the number of target photons for $\gamma$--$\gamma$
collisions, inducing a catastrophic cascade, and an overproduction of
X--ray with respect to what we observe (Svensson 1987).

For these reasons, we conclude that the required moderate pair
reprocessing is unlikely to occur in this source.

\section{Summary and Conclusions}

From the analysis of the {\it Beppo}SAX and ASCA data, and from the
construction of the simultaneous SED of the source at different epochs
we find:

\begin{itemize}
\item
{\it Beppo}SAX observed PKS 0528+134 in its faintest X--ray state,
characterized by the flattest X--ray spectrum, of energy index
$\alpha_x=0.49\pm 0.07$.

\item 
The three ASCA observations of March 1995, despite the different fluxes, 
are characterized by the same spectral index $\alpha_x=0.71\pm 0.04$.  
Fitting the sum of these data allows to determine with improved 
precision the absorbing hydrogen column density of the source, 
$N_{\rm H}=(5.3\pm 0.3)\times 10^{21}$ cm$^{-2}$, which corresponds 
to an optical absorption of $A_V=3\pm 0.17$.

% \item
% The, admittedly few, X--ray observations suggest a trend between X--ray 
% flux and slope, in the sense of steeper slopes for stronger fluxes.

% \item
% The opposite trend (flatter slope when the source is brighter)
% seems to be present in the $\gamma$--ray data.  
% The X--ray and $\gamma$--ray correlations together point to the presence 
% of a pivot in the spectrum, around $\sim$10 MeV.
% In this energy range the flux should vary less than in other high 
% energy bands.

\item
The overall simultaneous SED follows the general trend of blazars, as
found by Fossati et al. (1998).  
In fact the SED is characterized by two peaks, in the mid or far infrared 
and in the MeV band, respectively, as for the most luminous blazars.
Furthermore the $\gamma$--ray luminosity dominates the overall power
output during active states, again in agreement with the general
behavior of other powerful blazars (see e.g. von Montigny et al., 1995).

\item 
We have fitted the 1995 SED (high state) and the 1997 SED (low state)
with a homogeneous model, in which the emission from the far IR to the
UV is produced by the synchrotron process, and the high energy flux 
is due to the sum of the synchrotron self--Compton and the external 
Compton contributions. 
The former becomes more important (in the X--ray band) for increasing 
source power.
This accounts for the unusual observed  X--ray behavior 
(flatter state when the source was fainter).
Variations of the $\gamma$--ray flux and spectral shape might be 
explained by a flattening of the high energy part of the electron 
distribution.

\end{itemize}

\vskip 1 true cm
\noindent
{\bf Acknowledgments}

This research has made use of the NASA/IPAC extragalactic database
(NED) which is operated by the Jet Propulsion Laboratory, Caltech,
under contract with the National Aeronautics and Space Administration.
We have also used the public radio data of the Green Bank
Interferometer. 
We thank the italian space agency ASI for financial support
and the {\it Beppo}SAX space data center for constant help.
Annalisa Celotti and Giovanni Fossati acknowledge the
Italian MURST for financial support.

\end{document}